\documentclass[manuscript=true, review=false, anonymous=false, nonacm=true]{acmart}

\usepackage{enumitem}
\usepackage{microtype}
\usepackage{subfigure}
\usepackage{booktabs} 
\usepackage{hyperref}
\usepackage{times}
\usepackage{latexsym}
\RequirePackage{fix-cm}
\usepackage{amsmath, mathtools}
\usepackage{relsize}
\usepackage{footnote}
\usepackage[rightcaption]{sidecap}
\makesavenoteenv{tabular}
\makesavenoteenv{table}
\usepackage{multirow}
\usepackage{threeparttable}
\usepackage{array, url}
\usepackage{dblfloatfix}
\usepackage{graphicx}
\DeclareGraphicsExtensions{.pdf,.jpg,.png}
\usepackage{svg}
\svgpath{{../figures/}}
\graphicspath{{figures/}}
\usepackage{wrapfig}
\usepackage[framemethod=TikZ]{mdframed}

\mdfsetup{
    roundcorner=3pt,
    innerleftmargin=10pt,
    innerrightmargin=10pt,
    innertopmargin=5pt,
    innerbottommargin=10pt,
    frametitleaboveskip=5pt,
    skipabove=5pt,
    skipbelow=5pt
}

\mdfdefinestyle{greybox}{
    skipabove=10pt,
    skipbelow=20pt,
    middlelinecolor=gray!50,
    middlelinewidth=1.5pt,
    backgroundcolor=gray!5,
    frametitlefont={\normalfont\sffamily\bfseries}
}

\AtBeginDocument{%
  \providecommand\BibTeX{{%
    \normalfont B\kern-0.5em{\scshape i\kern-0.25em b}\kern-0.8em\TeX}}}

\setcopyright{acmcopyright}
\copyrightyear{2021}
\acmYear{2021}
\acmDOI{}

\acmConference[2021]{}{Special Issue on Accelerating AI on the Edge}{ACM TECS}
\acmBooktitle{Special Issue on Accelerating AI on the Edge}
\acmPrice{}
\acmISBN{}

\begin{document}

\title[SVEva Fair: Evaluating Fairness in Speaker Verification]{SVEva Fair: A Framework for Evaluating Fairness in Speaker Verification}

\author{Wiebke Toussaint}
\authornote{Corresponding author}
\email{w.toussaint@tudelft.nl}
\orcid{}
\author{Aaron Yi Ding}
\orcid{}
\affiliation{%
  \institution{Delft University of Technology}
  \country{The Netherlands}
}

\renewcommand{\shortauthors}{Toussaint and Ding}

\begin{abstract}
Despite the success of deep neural networks (DNNs) in enabling on-device voice assistants, increasing evidence of bias and discrimination in machine learning is raising the urgency of investigating the fairness of these systems. Speaker verification is a form of biometric identification that gives access to voice assistants. Due to a lack of fairness metrics and evaluation frameworks that are appropriate for testing the fairness of speaker verification components, little is known about how model performance varies across subgroups, and what factors influence performance variation. To tackle this emerging challenge, we design and develop \textbf{SVEva}\footnote{SVEva stands for \textbf{S}peaker \textbf{V}erification \textbf{Eva}luation.}\textbf{ Fair}, an accessible, actionable and model-agnostic framework for evaluating the fairness of speaker verification components. The framework provides evaluation measures and visualisations to interrogate model performance across speaker subgroups and compare fairness between models. We demonstrate \textbf{SVEva Fair} in a case study with end-to-end DNNs trained on the VoxCeleb datasets to reveal potential bias in existing embedded speech recognition systems based on the demographic attributes of speakers. Our evaluation shows that publicly accessible benchmark models are not fair and consistently produce worse predictions for some nationalities, and for female speakers of most nationalities. To pave the way for fair and reliable embedded speaker verification, \textbf{SVEva Fair} has been implemented as an open-source python library and can be integrated into the embedded ML development pipeline to facilitate developers and researchers in troubleshooting unreliable speaker verification performance, and selecting high impact approaches for mitigating fairness challenges.

\end{abstract}



\maketitle

\section{Introduction}
\label{introduction}

Today, over 4 billion voice assistants are deployed on mobile phones and smart devices. This number is estimated to double in the next 3 years. Speaker verification components offer voice-based biometric identification in automated speech recognition systems on mobile phones and smart speakers. In recent years, deep neural networks (DNN) have become the state-of-the-art approach for developing speaker verification components \cite{snyder2017deep, li2017deep, snyder2018xvectors}. A key advantage of DNNs is that they can be trained in an end-to-end fashion using only speaker labels \cite{heigold2016endtoend}. End-to-end DNN models greatly simplify model training and inference, which makes them particularly attractive for embedded applications. Running speaker verification components on-device is desirable, as this reduces latency and privacy concerns associated with sending sensitive data to the cloud. Considerable research efforts have thus been invested into developing speaker verification components for low-resource devices that can be deployed with streaming data input in real-time \cite{he2019streaming}. Pre-trained speaker verification DNN models are now publicly available \cite{heo2020clova} and can be accessed easily by developers to build new voice-based applications that incorporate speaker verification.

Despite the commercial success of on-device voice assistants and speaker verification, automated speech recognition systems are increasingly scrutinised for being biased \cite{koenecke2020racial}. These investigations follow trends in the broader machine learning (ML) community, which is uncovering increasing forms of bias and discrimination in machine learning systems~\cite{mehrabi2019survey}. Fairness has thus become an important consideration in the development of machine learning applications and in the framing of ethical artificial intelligence \cite{mittelstadt2016ethics}. In the speech recognition community it is well known that speaker demographics such as age, accent and gender affect the performance of speaker verification \cite{hansen2015speaker}. Speaker and technology variability is amplified in embedded applications, which are context dependent, operate on heterogeneous devices, and cater to very diverse populations of end-users. It is thus surprising that fairness is under-researched in the speaker verification community. Currently, speaker verification models and frameworks do not measure model performance across speaker subgroups. Consequently, little is known about how model performance varies across subgroups, and what factors influence performance variation. To ensure reliable performance in embedded systems, deeper evaluation of speaker verification components is needed to characterise fairness. In particular, tools that test the consistency of model performance across demographic speaker subgroups could support the evaluation of fairness of speaker verification components.

This paper contributes an accessible, actionable and model-agnostic framework for evaluating the fairness of speaker verification components. The framework provides evaluation measures and visualisations to interrogate model performance across speaker subgroups and compare fairness between models. The objective of doing this is to support developers and researchers in troubleshooting unreliable speaker verification performance, and selecting high impact approaches for mitigating fairness challenges. The framework is aligned with speaker verification evaluation best practice and aims to be compliant with the EU legal framing of non-discrimination. It has been implemented and open-sourced as a python library, \textbf{SVEva Fair}. We demonstrate the potential of \textbf{SVEva Fair} in an in-depth fairness analysis of the open-source speaker verification benchmark VoxCeleb Trainer, trained on the popular VoxCeleb datasets. This paper is one of the first research studies to investigate fairness considerations in embedded machine learning (ML) applications and contributes to the growing body of work on testing methodologies for embedded ML in the IoT context. Our findings shall inspire future research towards fair and reliable embedded speaker verification, and the responsible development of Edge AI.

The paper is structured as follows. In Section 2. we start by providing a background on speaker verification and its evaluation, review related work on fairness in ML and on fair speaker verification. We then present an overview of \textbf{SVEva Fair}, our proposed framework for evaluating fairness of speaker verification components in Section 3. In Section 4 we introduce a case study in which we use \textbf{SVEva Fair} to evaluate the fairness of pre-trained models released with the VoxCeleb Trainer benchmark. We present the evaluation of the case study in Section 5, and a detailed discussion of results, insights and limitations in Section 6. Finally, in Section 7 we highlight directions for future work and conclude.
\section{Background and Related Work}
\label{literature}

\subsection{Overview of Speaker Verification}

Speaker verification tasks are classified as text-dependent if the spoken phrases are fixed or text-independent if not, prompted if text is read or spontaneous if not \cite{greenberg2020two}. Spontaneous text-independent speaker verification is the most general task and the one that we investigate in this study. The traditional speaker verification protocol consists of three stages: model training, speaker enrollment and evaluation of speaker pairs. End-to-end speaker verification with DNNs combines all three stages in a single model \cite{heigold2016endtoend}. Early DNN-based approaches extracted features from utterances of enrolled speakers to generate speaker embeddings, and then optimised an objective function of the distance between same speaker and different speaker embedding pairs \cite{snyder2017deep}. During evaluation, utterances from enrolled and test speakers were then scored by the distance metric used in the objective function. Current state-of-the-art approaches use convolutional neural networks, which have been very successful in computer vision tasks, to directly learn speaker embeddings from audio spectrograms \cite{Nagrani2017}. These architectures are trained on paired embeddings, and aim to minimise the distance between embeddings from the same speaker, while maximising the distance between embeddings of different speakers. During inference, the DNN model outputs a distance-based score between the enrolled and test speaker embeddings, which corresponds directly to speaker similarity.

\subsection{Speaker Verification Evaluation}

\begin{figure*}[hbt]
    \centering
    \includegraphics[width=\textwidth]{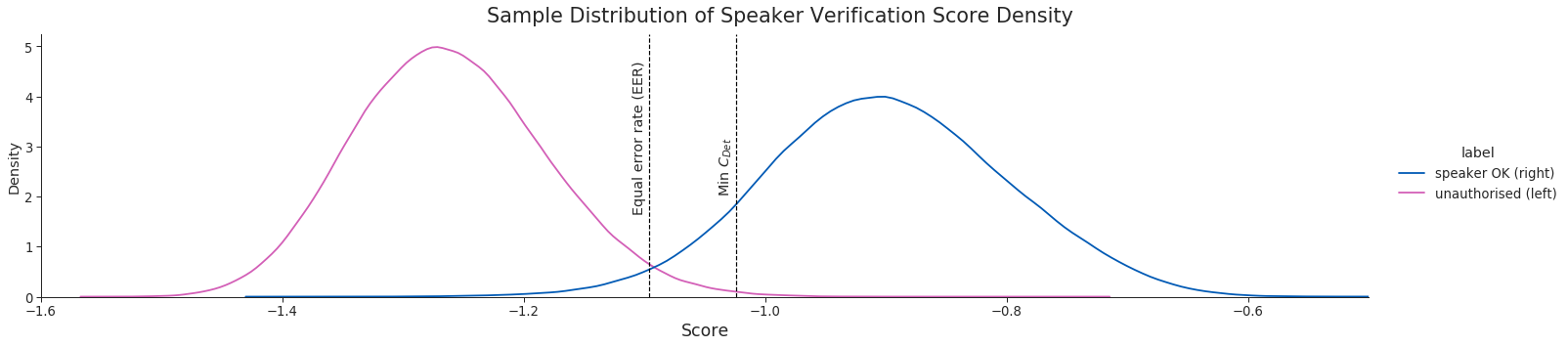}
    \caption{Distribution of speaker verification scores: the right distribution are trials with a positive label (speaker authorised), the left distribution are trials with a negative label (speaker unauthorised).}
    \label{fig:scores_example}
\end{figure*}

Figure \ref{fig:scores_example} shows an example of the output score distribution of an end-to-end DNN based speaker verification model. The performance of a speaker verification component is determined by its false positive rate (FPR) and false negative rate (FNR) at a particular threshold value to which the component has been tuned \cite{greenberg2020two}. Speaker embeddings with a score value to the left of the threshold are classified as unauthorised, while embeddings with scores to the right of the threshold are classified as authorised. As the two distributions overlap, classification is not perfect. At a selected threshold value there will be false positives, i.e. unauthorised speakers with a score value to the right of the threshold, and false negatives, i.e. authorised speakers with a score to the left of the threshold. The two error rates are influenced by the size of the overlapping area, as well as the shape, the skew and the kurtosis of the distributions. The dotted lines in Figure \ref{fig:scores_example} are popular thresholds at which speaker verification models are evaluated. The equal error rate (EER) is the threshold at which the FPR and FNR are the same. At the $min\ C_{Det}$ threshold the detection cost function (see Eq. \ref{eq:cdet}) is minimised. 

\begin{SCfigure*}[15][h]
    \centering
    \caption{Detection Error Trade-off (DET) Curve of a speaker verification model: the blue line shows false positive and false negative error rates at different score values. For example, at the blue triangle the score = -1.024, FPR = 0.27\% and FNR = 10.36\%}
    \includegraphics[width=0.4\textwidth]{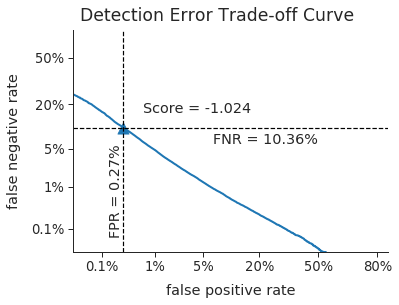}
    \label{fig:det_curve_example}
\end{SCfigure*}

In embedded speaker verification systems the FNR can reduce the usability and safety of the system, while the FPR can compromise security and privacy. It is accepted that the two error rates present a trade-off. Selecting an appropriate threshold is considered an application-specific design decision \cite{NIST2020}. Effective visualisations can be used to analyse the trade-off and consider system performance across various thresholds. Detection Error Trade-off (DET) curves as shown in Figure \ref{fig:det_curve_example} visualise the FPR and FNR at different operating thresholds on the x- and y-axis of a normal deviate scale \cite{Martin1997Det}. They are the recommended approach for visualising speaker verification model performance \cite{greenberg2020two}.

\begin{table}[b]
    \small
    \centering
    \renewcommand{\arraystretch}{1.2}
    \begin{tabular}{l|l|c|c}
        \textbf{Name} & \textbf{Organiser} & \textbf{Years} & \textbf{Metrics} \\
        NIST SRE \cite{greenberg2020two} & US National Inst. of Standards \& Tech. & 1996 - 2021 & Detection Cost Function\\
        Speakers in the Wild SRC \cite{mclaren2016speakers} & at Interspeech 2016 & 2016 & $min\ C_{Det}$* (SRE2016), $R_{prec}$, $C_{llr}$\\
        VoxCeleb SRC \cite{Nagrani2020Voxsrc} & Oxford Visual Geometry Group & 2019 - 2021 &  $min\ C_{Det}$* (SRE2018), EER\\
        Far-Field SVC \cite{qin2020interspeech} & at Interspeech 2020 & 2020 & $min\ C_{Det}$*, EER \\
        Short Duration SVC \cite{zeinali2019shortduration} & at Interspeech 2021 & 2020 - 2021 & $norm\ min\ C_{Det}$* (SRE08)\\
        SUPERB benchmark \cite{yang2021superb} & CMU, JHU, MIT, NTU, Facebook AI & 2021 & EER* \\
        \end{tabular} {\medskip}
    \caption{Evaluation metrics for Speaker Verification and Recognition Challenges (SVC and SRC) (* denotes primary metric)}
    \label{tab:sv_challenges}
\end{table}

Speaker recognition challenges have played an important role in evaluating and benchmarking advances in speaker verification techniques. They were first initiated within the Information Technology Laboratory of the US National Institute of Standards and Technology (NIST) to conduct evaluation driven research on automated speech recognition \cite{greenberg2020two}. The NIST Speaker Recognition Evaluation (SRE) challenges and their associated evaluation plans have been important drivers of speaker verification evaluation, and remain the dominant guideline in the field. Table \ref{tab:sv_challenges} summarises recent challenges, their organisers and the metrics used for evaluation. The NIST SREs recommend the use of the detection cost function, a weighted sum of FPRs and FNRs, for evaluating speaker verification performance. Most challenges have adopted the minimum of the detection cost function, $min\ C_{Det}$, as their primary metric. However, the NIST SREs have modified this function over time, and different challenges use different versions of the metric. In this study we adopt the detection cost function in Equation \ref{eq:cdet} from the NIST SRE 2019 Evaluation Plan \cite{NIST2019}.

\begin{equation}
\begin{split}
    C_{Det}\left(\theta\right) & = C_{FN} \times P_{Target} \times P_{FN}\left(\theta\right) + C_{FP} \times \left(1 - P_{Target}\right) \times P_{FP}\left(\theta\right) \\
    & P_{Target} = 0.05, \;
    C_{FN} = 1, \;
    C_{FP} = 1    
\label{eq:cdet}
\end{split}
\end{equation}

Even though the EER is a popular error metric in many of the challenges, the NIST SREs do not promote its use for speaker verification evaluation \cite{greenberg2020two}, as it cannot weight false positives and false negatives differently. Yet, most applications strongly favour either a low FPR or a low FNR. With the exception of some historic NIST evaluations that have considered speaker verification performance for particular demographic groups, none of the challenges consider fairness.

\subsection{Fairness in Machine Learning}

Fairness in machine learning (ML) has been studied extensively over the past decade \cite{mehrabi2019survey}. Fairness issues constitute a discriminatory action, typically against an individual or a group of people with one or more protected attributes \cite{mehrabi2019survey}. Protected attributes can be location and context dependent, and are often defined by law such that discrimination based on these attributes is illegal. EU non-discrimination law, for example, prohibits both direct and indirect discrimination based on race and ethnicity, gender, religion and belief, age, disability, or sexual orientation \cite{wachter2021bias}. Indirect discrimination happens when an attribute that seems to be independent of protected attributes is used in decision making such that it inadvertently disadvantages a protected group. With increasing global-scale commercial deployments of ML products, investigating and evaluating fairness of ML products has significant evidence in the literature and is a matter of urgency~\cite{buolamwini2018gendershades}.

Numerous metrics have been proposed to evaluate fairness of ML systems \cite{verma2018fairness}. Fairness metrics can be categorised as individual, group or subgroup fairness \cite{mehrabi2019survey}. Group fairness metrics treat different groups of people equally, which aligns with our objective of evaluating the performance of speaker verification components across demographic subgroups. However, not all metrics evaluate fairness on equal grounds. Metrics may thus lead to different outcomes when judging the fairness of ML systems, which makes the selection of fairness metrics a normative decision \cite{wachter2021bias}. The choice of metric matters less when it is used for diagnostic and testing purposes, and when ground-truth labels can be known exactly. This reduces the constraints on selecting an appropriate metric for evaluating the fairness of speaker verification components during the development process, as speaker labels are always exactly known. Four common metrics that evaluate group fairness are demographic parity, conditional statistical parity, equal opportunity and equalised odds \cite{mehrabi2019survey}. The first three of these metrics only consider the fairness based on positive outcomes, or the true positive rate. Equalised odds, on the other hand, requires protected and unprotected groups to have equal true and false positive rates. Mathematically this is equivalent to equal FNR and FPR across groups \cite{verma2018fairness}. Speaker verification components must trade-off FPR and FNR, which makes the equalised odds metric the most appropriate fairness definition for our application.

\subsection{Fair Speaker Verification}

It is well known that automated speech recognition is sensitive to demographic attributes of speakers \cite{hansen2015speaker}. In the past the effect of this has been investigated on telephone and broadcast corpora \cite{addadecker2005do}. More recently, studies have produced evidence that commercial automated caption systems have a higher word error rate for speakers of colour \cite{tatman2017effects}. Similar racial disparities exist in commercial speech-to-text systems, which are strongly influenced by pronunciation and dialect \cite{koenecke2020racial}. In the speaker verification domain, research on fairness is scarce. Fenu et. al propose a benchmark to evaluate the fairness of end-to-end deep learning models with Thin-ResNet and X-vector architectures \cite{fenu2020improving}. The study trains several models of young and old, female and male speakers in English and Spanish using the Mozilla Common Voice dataset. The benchmark is limited in that it only considers the EER metric, and the fairness evaluation appears to be done in a manual and adhoc manner as no fairness metric has been defined. This highlights the need for a rigorous speaker verification evaluation framework to test fairness in a reliable manner.

\begin{mdframed}[style=greybox, frametitle={Take-away}]
    End-to-end speaker verification with deep neural networks has delivered state-of-the art results for speaker verification in modern speech recognition systems. Today, these speaker verification components are deployed on billions of commercially-available consumer products, like voice assistants on mobile phones and smart speakers. Yet, even though it is well known that deep neural networks often produce biased and discriminatory predictions, issues of fairness are currently not considered in the evaluation of speaker verification models. \textbf{SVEva Fair} aims to address this gap with a framework for evaluating the fairness of speaker verification components.
\end{mdframed}
\section{SVEva Fair}
\label{framework}

This section describes the \textbf{SVEva Fair} evaluation framework. The objective of the framework is to evaluate the fairness of speaker verification components across demographic subgroups in the development of edge intelligence applications. Fairness is currently not covered by other frameworks that evaluate speaker verification or embedded ML systems. With \textbf{SVEva Fair} we aim to equip application developers of speaker verification components with a tool to interrogate two questions:
\begin{enumerate}
    \item Fairness: Does the performance of speaker verification components vary across speaker subgroups for a particular model?
    \item Comparison: How does fairness compare across speaker verification models?
\end{enumerate}
\textbf{SVEva Fair} is intended to be used as a domain-specific diagnostic tool that can assist developers in testing the fairness of speaker verification components, while facilitating the development workflow. With this in mind, \textbf{SVEva Fair} has been developed to satisfy the following design principles:
\begin{itemize}
    \item model and inference workflow \textbf{agnostic}
    \item aligned with \textbf{best practice} for speaker verification evaluation
    \item \textbf{accessible} to developers
    \item \textbf{actionable}, supporting informed decision-making for developing and deploying fair speaker verification
    \item \textbf{compliant} with the EU legal framing of non-discrimination
\end{itemize}

\noindent
Next we present the evaluation measures developed for \textbf{SVEva Fair}, and provide an overview of the evaluation workflow.

\subsection{Evaluation Measures}
Speaker verification components are used as a form of biometric identification in embedded systems and mobile devices. They thus function as gate keepers that filter out intruders who are not authorised to access the services enabled by a product. The performance of a speaker verification component for a particular application is determined by its false positive and false negative rate at the threshold value to which the component has been tuned. In a consumer product context we define speaker verification as fair if the component works equally well for all users, that is, if performance does not depend on a user's protected attributes such as age, sex or race. The detection cost $C_{Det}$ (see eq. \ref{eq:cdet}), which is used to evaluate speaker verification models, is a weighted sum of false positive and false negative error rates at a particular threshold value. Weights should be determined based on the requirements of the application. $C_{Det}$ can be viewed as a weighted equalised odds ratio, which in its unweighted form is a popular fairness metric. We thus use $C_{Det}$ as a proxy for fairness. In a fair speaker verification component, the $C_{Det}$ values of individual subgroups should lie close to each other and close to the overall $C_{Det}$ value for all subgroups. 

\subsubsection{Subgroup $C_{Det}$ Ratios}\hfill\\
\label{cdetratios}
\textbf{SVEva Fair} treats fairness as a relative measure that is determined by comparing the speaker verification performance of a subgroup against a baseline. Given a test dataset for evaluation, we use the overall test set performance as baseline against which we compare the performance for all subgroups. For each subgroup we evaluate the performance at two threshold values, the overall minimum detection cost of the test set, $C_{Det}\left(\theta_{@\ overall\ min}\right)$, and the subgroup minimum $C_{Det}\left(\theta_{@\ SG\ min}\right)$. For a fair speaker verification component the performance of all subgroups should lie within an acceptable range of the baseline's overall minimum detection cost. We quantify the relative fairness for each subgroup with the ratio of the subgroup $C_{Det}\left(\theta\right)^{SG}$ to the overall $C_{Det}\left(\theta\right)^{overall}$ at the overall minimum threshold value. If the ratio is greater than 1, the subgroup performance is worse than the overall performance, and the speaker verification component is not fair for that subgroup.

\begin{equation}
    {C_{Det}\ ratio\ overall_{min}}^{SG} = \frac{C_{Det}\left(\theta_{@\ overall\ min}\right)^{SG}}{C_{Det}\left(\theta_{@\ overall\ min}\right)^{overall}}
\end{equation}

\noindent
For optimal subgroup performance we would also expect that the subgroup minimum $C_{Det}\left(\theta_{@\ SG\ min}\right)^{SG}$ lies close to the subgroup detection cost at the overall minimum, $C_{Det}\left(\theta_{@\ overall\ min}\right)^{SG}$. We calculate the ratio between the two detection costs to determine if there is potential performance gain from tuning thresholds to individual subgroups. If the ratio is less than 1, then the subgroup performance will benefit from being tuned to the threshold at its own minimum, rather than the overall minimum $C_{Det}$.
\begin{equation}
    {C_{Det}\ ratio\ SG_{min}}^{SG} = \frac{C_{Det}\left(\theta_{@\ SG\ min}\right)^{SG}}{C_{Det}\left(\theta_{@\ overall\ min}\right)^{SG}}
\end{equation}

\noindent
While the subgroup $C_{Det}$ ratios are a useful measure for analysing intra-model fairness, they are specific to their subgroups and cannot be used to compare fairness across models. 

\subsubsection{Fairness Index from $C_{Det}$ Ratios}\hfill\\
We define a Fairness Index to interrogate which model reduces the performance of subgroups the least, and thus compare fairness across models. The Fairness Index is calculated from the $C_{Det}$ ratios to quantify the total performance reduction of a model across all subgroups ($SG$). Only those subgroups that experience a performance reduction contribute to the index, as this has negative real-life consequences. The closer the index is to 0, the fairer the model, meaning the lower the difference between the overall model performance and subgroup performance. The Fairness Index is an dimensionless measure that quantifies the relative performance reduction of a model across subgroups. It can thus be used to compare fairness across models.

\begin{equation}
    Fairness\ Index = \sum_{SG} {C_{Det}\ ratio\ overall_{min}}^{SG} - 1 \left[{C_{Det}\ ratio\ overall_{min}}^{SG} > 1\right]
\label{eq:fairnessindex}
\end{equation}

\subsubsection{Visualisation}\hfill\\
\textbf{SVEva Fair} supports three types of visualisation: DET curves, score distributions and $C_{Det}$ ratios. The DET curves visualise the possible performance range for individual subgroups, or a selection of subgroups and models. A baseline DET curve can be added to compare subgroup DETs to the overall DET, and various threshold values can be plotted: the equal error rate and minimum $C_{Det}$ threshold values, optimised for overall performance and for each subgroup. The DET curves enable model comparison at a glance, and give a first impression of the subgroups for which the model is not fair. While the DET curves are a translation of model outputs into false positive and false negative rates, the score distributions give deeper insights into the actual model output. Visualising how predictions are distributed can be helpful to gain an understanding of the limits of the model, and thus determine  mitigation strategies for fairness challenges. Like the DET curves, the score distributions can be visualised for individual subgroups or compared across subgroups and models. Finally, the plot of $C_{Det}$ ratios focuses on fairness only, and visualises the ${C_{Det}\ ratio\ overall_{min}}^{SG}$ of all subgroups across two models. This highlights variations in subgroup performance within a model and facilitates the comparison of subgroup performance across the models.   

\subsection{Overview of Evaluation Framework}

\begin{figure*}[hbt]
    \centering
    \includegraphics[width=0.7\textwidth]{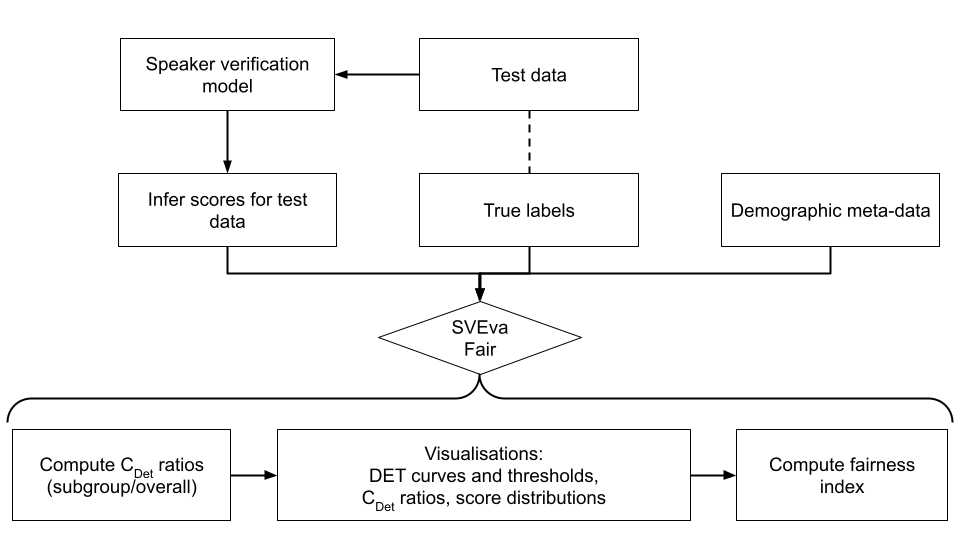}
    \caption{\textbf{SVEva Fair} Framework for evaluating fairness of speaker verification components}
    \label{fig:sveva_framework}
\end{figure*}

Figure \ref{fig:sveva_framework} shows the \textbf{SVEva Fair} evaluation framework. The framework has been implemented as an open-source python library\footnote{https://github.com/wiebket/sveva-fair}. The inputs to the framework are a speaker verification model, labelled test data, and demographic meta-data. The speaker verification model is treated as a black box, as long as the output is a distribution of numerical scores. The model is used to infer scores for the test data. These scores are then submitted to SVEva Fair, together with the true labels of the test data and associated speaker meta-data for constructing relevant subgroups. \textbf{SVEva Fair} is then used to compute $C_{Det}$ ratios, to visualise DET curves, score distributions and $C_{Det}$ ratios across models, and to compute the Fairness Index. The outputs of the framework can be used to evaluate the fairness of individual speaker verification models and to compare models. These insights are important inputs for determining strategies to improve the fairness of speaker verification components. Next we present a case study using \textbf{SVEva Fair}.

\section{Case Study with VoxCeleb Trainer}
\label{casestudy}

In this section we demonstrate the use of \textbf{SVEva Fair} by evaluating the fairness of models released with VoxCeleb Trainer~\cite{heo2020clova}, an open-source speaker verification training framework. VoxCeleb Trainer uses the popular VoxCeleb datasets~\cite{Nagrani2020a}, which are also used in other speaker verification benchmarks~\cite{yang2021superb} and challenges~\cite{NIST2020}\cite{Nagrani2020Voxsrc}. At the time of writing, VoxCeleb Trainer is recommended as the unofficial baseline code for speaker verification using the VoxCeleb datasets\footnote{https://www.robots.ox.ac.uk/~vgg/data/voxceleb/vox2.html}. \textbf{SVEva Fair} is one of the first comprehensive studies looking at fairness of speaker verification models trained with the VoxCeleb datasets.

\subsection{The VoxCeleb Datasets}
The VoxCeleb datasets~\cite{Nagrani2020a}, VoxCeleb1 and VoxCeleb2, contain short clips of audio-visual data of human speech, extracted from YouTube interviews with celebrities. VoxCeleb1 contains over 100,000 utterances by 1,251 speakers, with metadata for speaker sex and nationality. VoxCeleb2 contains over 1 million utterances by 6,112 speakers and metadata for speaker sex. The datasets are split into a training set and a test set. Each training set is disjoint from its test set and the other training set. VoxCeleb1 can thus be used as test set for models trained on VoxCeleb2. To support using VoxCeleb1 for testing, two additional test sets have been defined from speaker pairs in VoxCeleb1. VoxCeleb1-E consists of 581,480 speaker pairs covering all 1251 speakers in VoxCeleb1, sampled at random. VoxCeleb1-H consists of 1190 speakers, combined into 552,536 speaker pairs with the same sex and nationality. Speakers have only been included if there are at least 5 unique speakers with the same sex and nationality. 

\begin{table}[hbt]
\centering
\begin{tabular}{lcc}
\textbf{Nationality}       & \textbf{Sex} & \textbf{Unique speakers} \\ \noalign{\smallskip}
\multirow{2}{*}{USA}       & f               & \textbf{368}      \\
                           & m               & \textbf{431}      \\
\multirow{2}{*}{UK}        & f               & \textbf{88}       \\
                           & m               & \textbf{127}      \\
\multirow{2}{*}{Canada}    & f               & \textbf{25}       \\
                           & m               & \textbf{29}       \\
\multirow{2}{*}{Australia} & f               & \textbf{12}       \\
                           & m               & \textbf{25}       \\
\multirow{2}{*}{India}     & f               & \textbf{11}       \\
                           & m               & \textbf{15}       \\
\multirow{2}{*}{Norway}    & f               & \textbf{7}        \\
                           & m               & \textbf{13}       \\
\multirow{2}{*}{Ireland}   & f               & \textbf{5}        \\
                           & m               & \textbf{13}       \\
New Zealand                & m               & \textbf{6}        \\
Germany                    & f               & \textbf{5}        \\
Italy                      & f               & \textbf{5}        \\
Mexico                     & m               & \textbf{5}       
\end{tabular}{\smallskip}
\caption{VoxCeleb1-H speaker distribution across nationality and sex subgroups.}
\label{tab:voxceleb1h_demographics}
\end{table}

For embedded voice assistants, the sex-nationality pairs of VoxCeleb1-H present the most realistic evaluation condition. Embedded applications are context and location dependent. In homes, cars, offices or public spaces, where embedded voice assistants are used or accidentally triggered most frequently, speakers oftentimes have the same nationality, speak the same language with similar accents and are of the same sex. For example, a mother may sound very similar to her teenage daughter, or two colleagues in Delhi, India are likely to speak with the same accent. We thus use VoxCeleb1-H as test data in this case study. The demographic speaker distribution across nationality and sex subgroups in VoxCeleb1-H is shown in Table \ref{tab:voxceleb1h_demographics}. 44\% of the speakers are female. US nationals make up 67\% of the dataset and are the most represented nationality. The largest subgroup, US males, contributes 36\% of speakers, while the four smallest subgroups collectively make up less than 2\% of speakers in the dataset. VoxCeleb2 has 61\% male speakers. Again, the most represented nationality are US speakers, which make up 29\% of speakers. Other speaker nationalities are difficult to discern from the available documents and metadata. 

\subsection{Setting up VoxCeleb Trainer for Evaluation with SVEva Fair}
The VoxCeleb Trainer makes two pretrained baseline models available. ResNetSE34V2 is described in \cite{heo2020clova}, where it is called the performance optimised model, H/ASP. ResNetSE34L is described in \cite{chung2020defence}, where it is called Fast ResNet-34. We present a brief overview of the models and training procedures. Both models are based on a 34-layer ResNet trunk architecture, and have been trained on the 5994 speakers from the training set of VoxCeleb2. ResNetSE34V2 has been trained with data augmentation, while ResNetSE34L has not. Main differences in the model architectures are that ResNetSE34L with 1.4 million parameters is considerably smaller than ResNetSE34V2 with 8 million parameters. The two models use different methods for aggregating frame-level features and different loss functions: ResNetSE34L uses self-attentive pooling and angular prototypical loss, while ResNetSE34V2 uses attentive statistical pooling and angular portotypical softmax loss. Finally, the input dimensions of the two models are different as ResNetSE34L has been optimised for fast execution and consequently has a smaller input and earlier stride. The stride at the first convolutional layer of ResNetSE34V2 has been removed.

\begin{figure*}[hbt]
    \centering
    \includegraphics[width=0.6\textwidth]{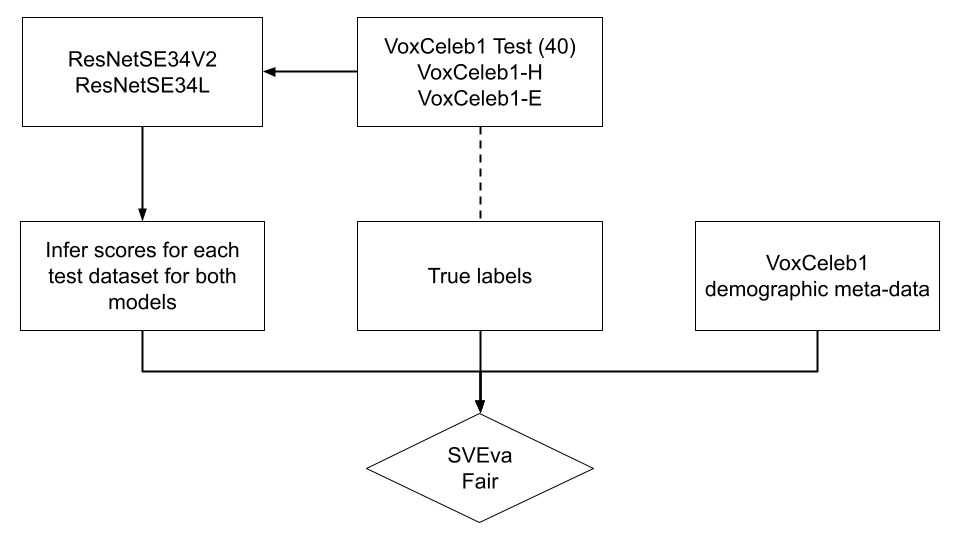}
    \caption{Experiment setup for \textbf{SVEva Fair} Case Study with VoxCeleb Trainer}
    \label{fig:casestudy_setup}
\end{figure*}

Our experiment setup for the case study is shown in Figure \ref{fig:casestudy_setup}. We downloaded the two models described above, and use them as black box predictors with the VoxCeleb Trainer inference pipeline in a Colab environment. We made minor modifications to the VoxCeleb Trainer code to speed up inference execution time, log evaluation results and reduce errors in data loading. Inference was done on the cleaned versions of the three VoxCeleb1 test sets described earlier. Using the VoxCeleb1 meta-data, we defined the demographic subgroups based on speaker sex and nationality. In the next section we present the \textbf{SVEva Fair} evaluation on the VoxCeleb1-H test data for this case study.
\section{Case Study Evaluation}
\label{evaluation}

The \textbf{SVEva Fair} evaluation framework aims to support inquiry into two questions: whether a particular speaker verification model is fair, and whether one speaker verification model is fairer than another. In this section we first investigate the fairness of the ResNetSE34V2 model described previously, and then compare the fairness of the ResNetSE34V2 and the smaller ResNetSE34L models.

\subsection{Performance of Speaker Verification Models Across Subgroups}

\begin{figure*}[t]
    \centering
    \includegraphics[width=\textwidth]{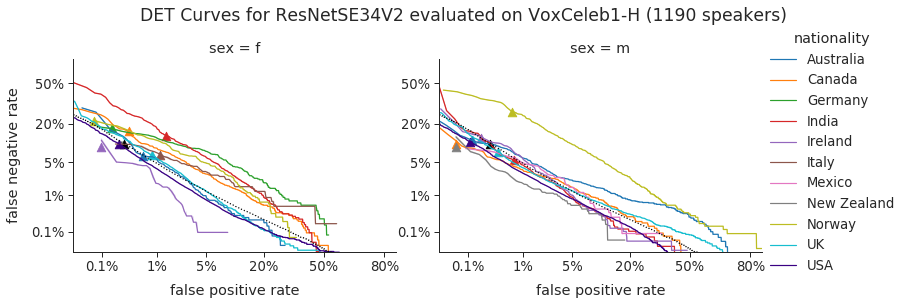}
    \caption{DET curves and subgroup thresholds at $C_{Det}(\theta_{@\ overall\ min})$ for ResNetSE34V2 evaluated on the VoxCeleb1-H test set. The dotted black lines and markers indicate the aggregate overall DET curve and threshold across all subgroups.}
    \label{fig:resnetse34v22_det_curves_sex}
\end{figure*}

We use \textbf{SVEva Fair} to interrogate whether the performance of the ResNetSE34V2 speaker verification model varies across speaker subgroups based on sex and nationality. Figure \ref{fig:resnetse34v22_det_curves_sex} shows DET curves for female and male speakers across 11 nationalities. By visually examining the plots, it is immediately evident that the DET curves of female speakers in the left column lie mostly above the dotted black DET curve that shows the aggregate overall performance. This is an early indicator that the model  is likely to perform worse than average for female speakers. Unsurprisingly, the false positive rates (FPR) and false negative rates (FNR) for most female speaker subgroups at the minimum overall threshold value are dispersed to the right and above the overall threshold. This indicates that the speaker verification component when tuned to its optimum overall operating threshold, works worse than aggregate for females. USA and Irish female speakers with test sample sizes of 368 and 5 speakers respectively, are an exception. The test sample size of Irish female speakers may be too small to conclude that the model truly performs better than the aggregate for this subgroup. For male speakers in the right column of Figure \ref{fig:resnetse34v22_det_curves_sex} most DET curves lie below the aggregate overall performance. The male subgroup FPRs and FNRs at the minimum overall threshold value lie below and to the left of the overall threshold. This indicates that male subgroups are likely to perform better than the aggregate. A noteable exception are Norwegian male speakers with a test sample size of 13 speakers. The model performs particularly bad for this subgroup.

\begin{figure*}[b]
    \centering
    \includegraphics[width=\textwidth]{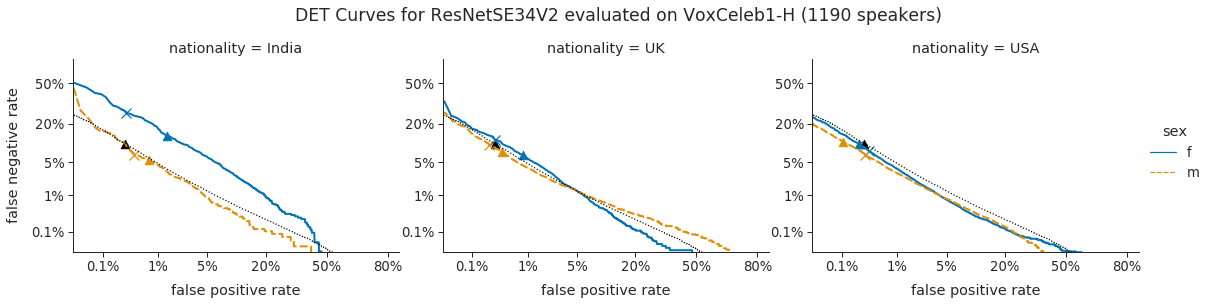}
    \caption{DET curves and thresholds for male and female speakers of Indian, UK and USA nationalities for ResNetSE34V2 evaluated on the VoxCeleb1-H test set. We use the following conventions: triangle markers represent the FPR and FNR at the overall minimum threshold $C_{Det}\left(\theta_{@\ overall\ min}\right)$, cross markers represent the FPR and FNR at the subgroup minimum threshold $C_{Det}\left(\theta_{@\ SG\ min}\right)$, and dotted black lines and markers are used for the overall DET curve and threshold.}
    \label{fig:resnetse34v22_det_curves_nationality}
\end{figure*}

Figure \ref{fig:resnetse34v22_det_curves_nationality} visualises the performance of ResNetSE34V2 for female and male speakers from India, the UK and the USA. The DET curve of female Indian speakers lies far above the overall aggregate, indicating that irrespective of the threshold, the model will always perform worse than aggregate for this subgroup. In the operating region around the tuned thresholds, the model also performs worse for female speakers from both the UK and the USA. Being tuned to $C_{Det}\left(\theta_{@\ overall\ min}\right)$ does not affect the FNR and improves the FPR of USA female and male speakers. For other speaker subgroups, especially UK females and Indian females and males, either the FPR or the FNR deteriorates significantly when tuned to the overall minimum. For all subgroups the threshold at the subgroup minimum, $C_{Det}\left(\theta_{@\ SG\ min}\right)$, shifts the FPR and FNR closer to those of the minimum overall threshold, suggesting that performance will improve when optimising thresholds for subgroups individually. The conclusions drawn from the visualisations are validated by the data presented in Table \ref{tab:cdet_ratios_resnetse34v22}. The table summarises the minimum $C_{Det}$ values optimised for overall and individual subgroup performance, and the two $C_{Det}$ ratios defined in Section \ref{cdetratios} for each subgroup. For our experiments $C_{Det}(\theta_{@\ overall\ min})^{overall} = 0.0077$ and this value was used to calculate the ${C_{Det}\ ratio\ overall_{min}}$.

\begin{table}[t]
\small
\centering
\begin{tabular}{lccccc}
\multicolumn{1}{l}{\textbf{Subgroup (SG)}} &
  \textbf{\begin{tabular}[c]{@{}c@{}}Unique \\ speakers\end{tabular}} &
  \textbf{\begin{tabular}[c]{@{}c@{}}$C_{Det}(\theta_{@\ overall\ min})$\end{tabular}} &
  \textbf{$C_{Det}(\theta_{@\ SG\ min})$} &
  \textbf{\begin{tabular}[c]{@{}c@{}}${C_{Det}\ ratio\ overall_{min}}$\end{tabular}} &
  \textbf{\begin{tabular}[c]{@{}c@{}}${C_{Det}\ ratio\ SG_{min}}$\end{tabular}} \\ \noalign{\smallskip}
\textbf{mexico\_m}     & 5   & 0.0045 & 0.0045 & 0.5768 & 1.0000 \\
\textbf{newzealand\_m} & 6   & 0.0052 & 0.0043 & 0.6668 & 0.8346 \\
\textbf{ireland\_f}    & 5   & 0.0055 & 0.0035 & 0.7109 & 0.6348 \\
\textbf{canada\_m}     & 29  & 0.0057 & 0.0052 & 0.7304 & 0.9146 \\
\textbf{usa\_m}        & 431 & 0.0065 & 0.0061 & 0.8357 & 0.9354 \\
\textbf{australia\_m}  & 25  & 0.0070 & 0.0068 & 0.9020 & 0.9713 \\
\textbf{usa\_f}        & 368 & 0.0071 & 0.0070 & 0.9224 & 0.9864 \\
\textbf{uk\_m}         & 127 & 0.0074 & 0.0070 & 0.9523 & 0.9492 \\ \cline{1-1} \cline{5-5} \noalign{\smallskip}
\textbf{ireland\_m}    & 13  & 0.0081 & 0.0080 & 1.0432 & 0.9842 \\
\textbf{australia\_f}  & 12  & 0.0089 & 0.0077 & 1.1523 & 0.8628 \\
\textbf{india\_m}      & 15  & 0.0095 & 0.0072 & 1.2200 & 0.7586 \\
\textbf{germany\_f}    & 5   & 0.0104 & 0.0092 & 1.3359 & 0.8885 \\
\textbf{canada\_f}     & 25  & 0.0112 & 0.0101 & 1.4501 & 0.8969 \\
\textbf{uk\_f}         & 88  & 0.0113 & 0.0086 & 1.4558 & 0.7613 \\
\textbf{norway\_f}     & 7   & 0.0114 & 0.0105 & 1.4711 & 0.9208 \\
\textbf{italy\_f}      & 5   & 0.0138 & 0.0052 & 1.7827 & 0.3777 \\
\textbf{norway\_m}     & 13  & 0.0199 & 0.0198 & 2.5720 & 0.9941 \\
\textbf{india\_f}      & 11  & 0.0200 & 0.0159 & 2.5766 & 0.7960
\end{tabular} {\medskip}
\caption{$C_{Det}$ values and ratios for subgroups at minimum overall and subgroup thresholds. Subgroups above the horizontal black line perform better than aggregate when tuned to the minimum overall threshold. The $C_{Det}\ ratio$ of these subgroups is less than 1 (second column from the left). The last column shows that all subgroups perform better when tuned to their own minimum.}
\label{tab:cdet_ratios_resnetse34v22}
\end{table}

\begin{figure*}[t]
    \centering
    \includegraphics[width=\textwidth]{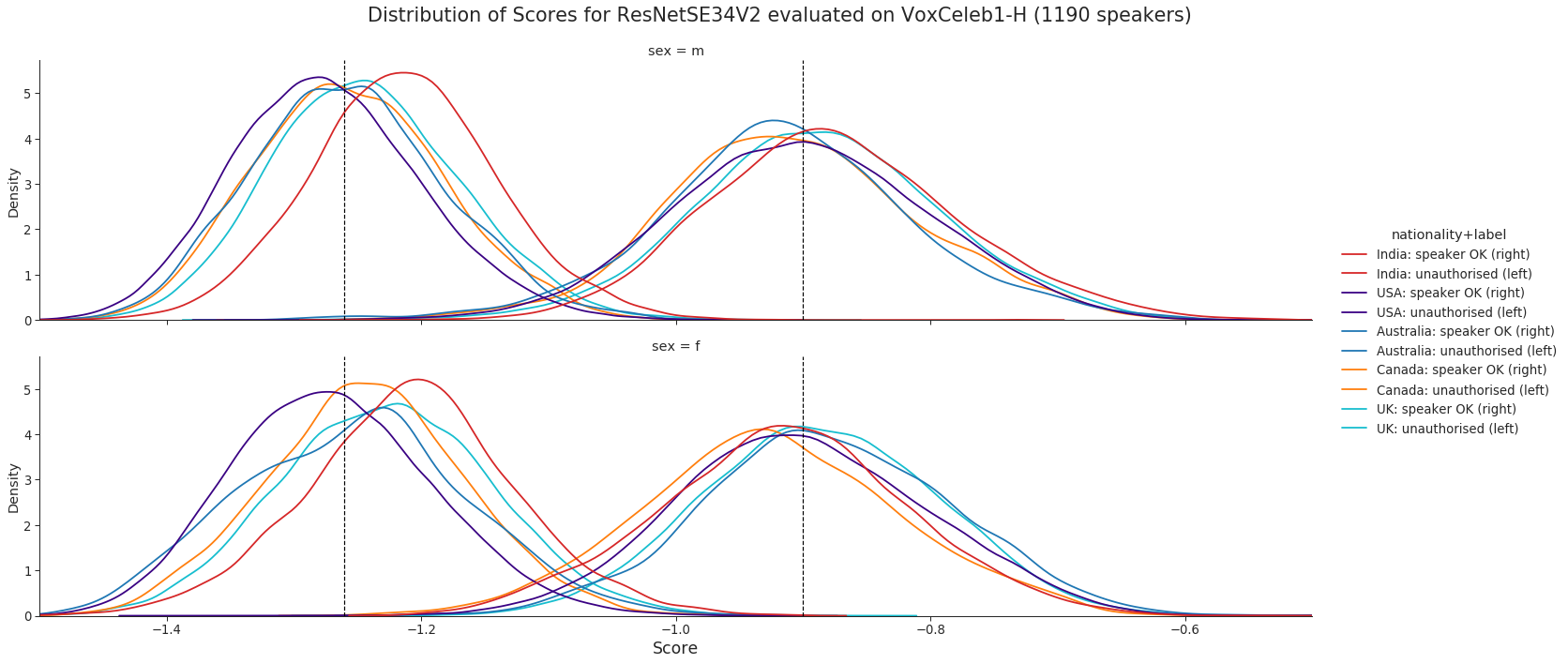}
    \caption{Distribution of scores by subgroup for ResNetSE34V2 evaluated on VoxCeleb1-H. The distribution on the right is for speakers tested against themselves and determines the FNR. The distribution on the left is for speakers tested against an unauthorised speaker and determines the FPR. The dotted black lines are the overall distribution means for same speaker and unauthorised speaker pairs across all subgroups.}
    \label{fig:scores_resnetse34v22}
\end{figure*}

Figure \ref{fig:scores_resnetse34v22} shows the score distributions that are generated by the model. For same speaker pairs that determine the FNR (right distribution), the score distributions lie close together for female and male speakers of all subgroups. However, female speakers have heavier left tails than males, indicating that the FNR for female speakers will be greater than for male speakers at a given FPR. For different speaker pairs (left distribution), the shape, mean, skewness, and kurtosis of the score distributions vary considerably across subgroups. The more right of the overall mean, the more right skewed and the heavier the right tail, the greater the FPR will be at a given FNR. When examining the intersection of the two distributions, it is also evident that subgroups have different intersection points and overlap areas. At a particular threshold score subgroups will thus have different FPR and FNR, which carries real-life consequences when speaker verification components are used in applications. Table \ref{tab:fpfn_ratios_resnetse34v22} shows the FPR and FNR ratios for subgroups at $C_{Det}(\theta_{@\ overall\ min})$. When tuned to this value, Indian female speakers have a FPR of 13.0387, indicating that the speaker verification component will grant access to an unauthorised speaker 13 times more frequently than average. On the other hand, male speakers from the USA have a FPR ratio of 1, which equals the overall FPR for all subgroups.

\begin{SCtable}[50][hbt]
\centering
\begin{tabular}{lccc}
\textbf{Subgroup} &
  \textbf{\begin{tabular}[c]{@{}c@{}}Unique \\ speakers\end{tabular}} &
  \textbf{\begin{tabular}[c]{@{}c@{}}FPR ratio \\ overall\end{tabular}} &
  \textbf{\begin{tabular}[c]{@{}c@{}}FNR ratio \\ overall\end{tabular}} \\ \noalign{\smallskip}
\textbf{mexico\_m}     & 5   & 0.0000  & 0.8173 \\
\textbf{canada\_m}     & 29  & 0.5171  & 0.9396 \\
\textbf{newzealand\_m} & 6   & 0.5218  & 0.8487 \\
\textbf{norway\_f}     & 7   & 0.6306  & 1.9682 \\
\textbf{ireland\_f}    & 5   & 0.9037  & 0.8408 \\
\textbf{usa\_m}        & 431 & 1.0000  & 1.0000 \\
\textbf{australia\_m}  & 25  & 1.1055  & 1.0745 \\
\textbf{germany\_f}    & 5   & 1.5023  & 1.6162 \\
\textbf{ireland\_m}    & 13  & 1.6864  & 1.1675 \\
\textbf{usa\_f}        & 368 & 2.0542  & 0.9287 \\
\textbf{canada\_f}     & 25  & 3.1483  & 1.4749 \\
\textbf{uk\_m}         & 127 & 3.5339  & 0.6986 \\
\textbf{australia\_f}  & 12  & 5.6031  & 0.6008 \\
\textbf{norway\_m}     & 13  & 6.0866  & 2.5233 \\
\textbf{india\_m}      & 15  & 6.6852  & 0.4975 \\
\textbf{uk\_f}         & 88  & 7.8514  & 0.6168 \\
\textbf{italy\_f}      & 5   & 10.3484 & 0.6202 \\
\textbf{india\_f}      & 11  & 13.0387 & 1.2497
\end{tabular}
\caption{FPR and FNR ratios for subgroups at $C_{Det}(\theta_{@\ overall\ min})$. The ratio is calculated by dividing the subgroup FPR and FNR by the overall FPR and FNR respectively. It thus presents a relative view on how much better or worse the subgroup error rates are in relation to the overall error rates.}
\label{tab:fpfn_ratios_resnetse34v22}
\end{SCtable}

\subsection{Comparing Speaker Verification Fairness Across Models}
\begin{figure*}[htb]
    \centering
    \includegraphics[width=\textwidth]{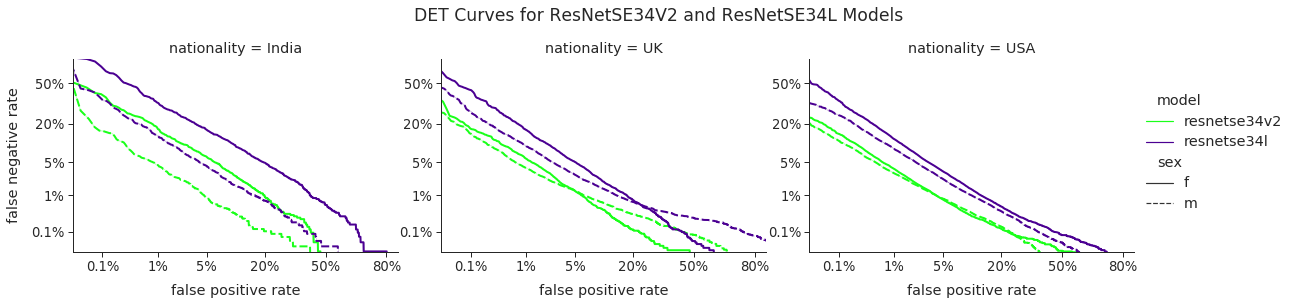}
    \caption{DET curves for female (solid) and male (dotted) speakers from India, the UK and the USA with ResNetSE34V2 (purple) and ResNetSE34L (green) models.}
    \label{fig:det_curves_models_nationality}
\end{figure*}

We now compare the fairness of the performance optimised ResNetSE34V2 model against the fairness of the smaller and speed optimised ResNetSE34L model. Figure \ref{fig:det_curves_models_nationality} shows the DET curves for female and male speakers from India, the UK and the USA for both models. As expected, all subgroup DET curves for ResNetSE34V2 (green) lie below those of ResNetSE34L (purple), confirming that the performance optimised model indeed has better performance. Surprisingly, the performance reduction does not affect all subgroups equally. For speakers from the UK and the USA the distance between the DET curves of females and males is greater for ResNetSE34L than ResNetSE34V2, indicating that female speaker will experience a greater performance degradation than male speakers when the speed optimised model is used. For Indian female speakers both models perform so poorly, that the ResNetSE34L DET curve for Indian males lies below the ResNetSE34V2 DET curve for Indian females. 

\begin{table}[b]
\small
\centering
\begin{tabular}{lcccc}
\multicolumn{1}{l}{\textbf{Subgroup}} &
  \textbf{\begin{tabular}[c]{@{}c@{}}Unique \\ speakers\end{tabular}} &
  \textbf{\begin{tabular}[c]{@{}c@{}}${C_{Det}\ ratio\ overall_{min}}$ \\ ResNetSE34V2\end{tabular}} &
  \textbf{\begin{tabular}[c]{@{}c@{}}${C_{Det}\ ratio\ overall_{min}}$ \\ ResNetSE34L\end{tabular}} &
  \textbf{\begin{tabular}[c]{@{}c@{}}${C_{Det}\ ratio\ overall_{min}}$ \\ difference\end{tabular}} \\ \noalign{\smallskip}
\textbf{india\_f}      & 11  & 2.5766 & 3.2869 & -0.7102 \\
\textbf{mexico\_m}     & 5   & 0.5768 & 1.2278 & -0.6510 \\
\textbf{germany\_f}    & 5   & 1.3359 & 1.5319 & -0.1959 \\
\textbf{norway\_f}     & 7   & 1.4711 & 1.6354 & -0.1643 \\
\textbf{canada\_m}     & 29  & 0.7304 & 0.8932 & -0.1628 \\
\textbf{australia\_m}  & 25  & 0.9020 & 1.0419 & -0.1398 \\
\textbf{usa\_f}        & 368 & 0.9224 & 0.9967 & -0.0743 \\ \cline{1-1} \cline{5-5} \noalign{\smallskip}
\textbf{usa\_m}        & 431 & 0.8357 & 0.8320 & 0.0037  \\
\textbf{india\_m}      & 15  & 1.2200 & 1.1657 & 0.0543  \\
\textbf{uk\_f}         & 88  & 1.4558 & 1.3566 & 0.0992  \\
\textbf{newzealand\_m} & 6   & 0.6668 & 0.5656 & 0.1012  \\
\textbf{ireland\_f}    & 5   & 0.7109 & 0.5952 & 0.1157  \\
\textbf{ireland\_m}    & 13  & 1.0432 & 0.9042 & 0.1390  \\
\textbf{canada\_f}     & 25  & 1.4501 & 1.3096 & 0.1404  \\
\textbf{uk\_m}         & 127 & 0.9523 & 0.8090 & 0.1433  \\
\textbf{australia\_f}  & 12  & 1.1523 & 0.9147 & 0.2376  \\
\textbf{italy\_f}      & 5   & 1.7827 & 1.4783 & 0.3044  \\
\textbf{norway\_m}     & 13  & 2.5720 & 2.1037 & 0.4683 
\end{tabular} {\medskip}
\caption{Comparision of ${C_{Det}\ ratio\ overall_{min}}$ for subgroups for ResNetSE34V2 and ResNetSE34L. The ratio difference is calculated by subtracting the ResNetSE34L ratio from the ResNetSE34V2 Ratio. A negative difference indicates that ResNetSE34V2 performs better, while a positive differences indicates that ResNetSE34L performs better.}
\label{tab:cdet_ratios_comparison}
\end{table}

\begin{figure*}[bt]
    \centering
    \includegraphics[width=0.7\textwidth]{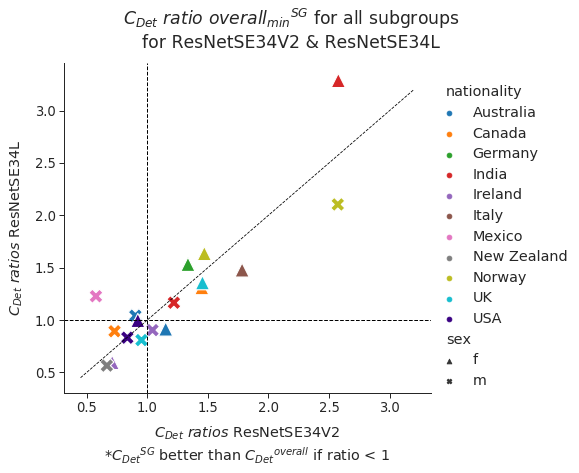}
    \caption{Subgroup ${C_{Det}\ ratio\ overall_{min}}$ for ResNetSE34V2 and ResNetSE34L. Subgroups in the bottom left perform best, those in the top right worst. Subgroup performance for ResNetSE34V2 and ResNetSE34L reduces when moving from left to right and bottom to top respectively. On the dotted black diagonal line subgroup performance is equivalent for the two models. ResNetSE34L performs better for subgroups below the line, while ResNetSE34V2 performs better above the line.}
    \label{fig:cdet_ratios_models}
\end{figure*}

Table \ref{tab:cdet_ratios_comparison} captures the ${C_{Det}\ ratio\ overall_{min}}$ for both models for all subgroups, and the difference between the ratios. From these values we calculated the Fairness Index (Equation \ref{eq:fairnessindex}) for ResNetSE34V2 as 16.06 and for ResNetSE34L as 16.14. The difference between the Fairness Indices of the two models is insignificant. Plotting the ${C_{Det}\ ratio\ overall_{min}}$ for all subgroups across both models enables us to closer analyse and compare fairness across the models. Figure \ref{fig:cdet_ratios_models} confirms that the trend in fairness challenges that we have observed with ResNetSE34V2 also applies to ResNetSE34L: the models perform better for male speakers (cross markers) than for female speakers (triangle markers), they perform particularly well for USA nationals and particularly poorly for Indian and Norwegian speakers. The plot also highlights that the smaller, speed optimised model does not impact fairness equally across subgroups. While fairness deteriorates for some subgroups, it improves for others. An example of this are Mexican male speakers, who outperform the aggregate with ResNetSE34V2 but perform worse than aggregate with ResNetSE34L. For UK female speakers the opposite is the case. Interestingly, there are few subgroups that experience a significant reduction in fairness, and more subgroups that experience a marginal improvement in fairness. This explains why the Fairness Indices of both models are similar.

\begin{mdframed}[style=greybox, frametitle={Take-away}]
        In this section we have demonstrated how \textbf{SVEva Fair} can be used to evaluate and compare the fairness of speaker verification components. Using the VoxCeleb Trainer benchmark, we show that available benchmark models are not fair and produce worse predictions for female speakers of most nationalities. Variation in predictive performance can also be observed across nationalities. The models perform particularly well for female and male speakers from the USA, and particularly poorly for female and male speakers from India and Norway. Surprisingly, when comparing fairness across models, we observe that it varies inconsistently. While fairness significantly deteriorates for some subgroups, it improves for others. This variation does not depend on sex, nationality, or subgroup sample size.
\end{mdframed}
\section{Insights and Discussion}

We have developed the \textbf{SVEva Fair} evaluation framework to equip developers of embedded speaker verification applications with a tool to assess and compare their fairness. In designing the framework, we have aimed to make \textbf{SVEva Fair} accessible, actionable, legally compliant, model and workflow agnostic, and aligned with evaluation best practices in the domain. The evaluation measures and perspective on fairness that we have chosen support these objectives. We have demonstrated how \textbf{SVEva Fair} can be used to evaluate the fairness of speaker verification components in a case study with the VoxCeleb Trainer benchmark. The analysis that we have presented is one of the first in-depth studies on the fairness of speaker verification. In this section we highlight insights that we have gained and their implication for real-life applications of embedded speaker verification. We then discuss integrating \textbf{SVEva Fair} into the embedded ML development pipeline, and point out limitations of the work.

\subsection{Insights on Fair Speaker Verification}

In the speaker verification domain detection error trade-off (DET) curves are used to analyse the performance of different models. With \textbf{SVEva Fair} we have shown that DET curves are also highly effective for analysing the performance of one or more models across speaker subgroups. They can thus be used as a tool to visualise and interrogate the fairness of speaker verification components. The detection cost function $C_{Det}(\theta)$, which is the recommended error function in the domain, is, by definition, a weighted sum of false positive rates (FPR) and false negative rates (FNR). This definition supports that of the equalised odds fairness metric, which requires FPR and FNR across subgroups to be equal. With \textbf{SVEva Fair} we propose that comparing the $C_{Det}$ values of subgroups at particular thresholds is a reasonable proxy for fairness. We define the $C_{Det}\ ratio\ overall_{min}$ as a metric that can be compared across subgroups to conveniently quantify the fairness of speaker verification components. To compare fairness across models, we suggest a Fairness Index to calculate the performance reduction of a model across subgroups from the $C_{Det}\ ratio\ overall_{min}$. 

Using the VoxCeleb Trainer speaker verification benchmark trained on the popular VoxCeleb dataset as a black box predictor, we show that two state-of-the-art ResNet-based speaker verification models are not fair. The predictive performance of ResNetSE3V2 varies considerably across subgroups. Male speakers from Canada, Australia and the USA experience a 10-25\% performance improvement in comparison to the average performance. For all speaker nationalities except Irish speakers, the model performs worse for female speakers than for male speakers. Performance degradation for female speakers ranges from 15\% to 257\% below average. Indian females experience the severest performance drop, with the model performing 2.6 times worse than average. To put this in perspective, this performance drop is over 60\% greater than the performance gain of using the best rather than the worst algorithm for training the VoxCeleb Trainer benchmark \cite{heo2020clova}. Most performance gains of algorithms thus pale in comparison to performance degradation due to fairness challenges. This is a motivation to consider fairness as an important research opportunity in embedded speaker verification. The fairness challenges that we observed with ResNetSE34V2 persist in ResNetSE34L. However, changes in fairness vary by subgroup, deteriorating for some subgroups and improving for others. This variation appears to be independent of sex, nationality, or subgroup sample size. 

\textbf{SVEva Fair} has given us insights into potential mitigation strategies and worthwhile research directions to improve fairness. Importantly, fairness depends on more than representative training data. As the $C_{Det}\ ratio\ SG_{min}$ shows, speaker verification will improve for all subgroups if they are tuned to their own threshold rather than the overall threshold. Developing algorithms that can dynamically select the optimal threshold for subgroups will improve the performance of speaker verification components. This is a challenging task, as subgroup membership is typically not known at run time. Analysis of the output score distributions shows that speaker verification is also highly dependent on this distribution. Further research is required to characterise the factors that affect this distribution and fairness across subgroups in speaker verification models, and subsequently to propose methods for improving speaker verification components.

\begin{mdframed}[style=greybox, frametitle={Take-away}]
    This study demonstrates that DET curves can be used to visualise the performance of models across subgroups. We motivate that $C_{Det}$ is a reasonable proxy for fairness that supports the definition of equalised odds. We use ratios of $C_{Det}$ values to quantify and compare fairness across subgroups, and show that two pre-trained speaker verification models trained on the VoxCeleb dataset are not fair. Drawing on our analysis, we highlight potential mitigation strategies that go beyond representative training data to improve fairness of speaker verification components.
\end{mdframed}

\subsection{Implications for Real-life, Embedded Voice Assistants}
We discuss two aspects of speaker verification evaluation that are of high importance in embedded applications: selecting an appropriate test set, and presenting a robust evaluation. Embedded speaker verification applications need to consider the demographic characteristics of their authorised and potential unauthorised users. In many smart home applications, for example mobile voice assistants or smart speakers, speaker verification components will need to distinguish between same-sex speakers of similar age, speaking similar languages with similar accents. A meaningful evaluation of speaker verification models needs to consider these typical scenarios. Of the 3 VoxCeleb1 test sets, only VoxCeleb1-H considers similar speakers. We found that the choice of test set has a significant and predictable impact on model performance. 

\begin{figure*}[hbt]
    \centering
    \includegraphics[width=0.5\textwidth]{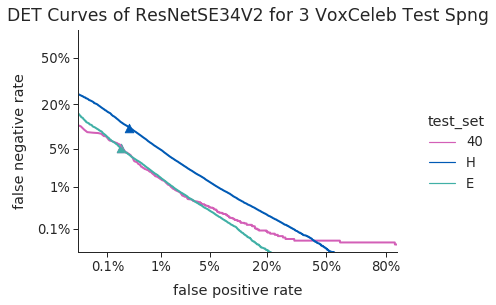}
    \caption{DET curves for 3 VoxCeleb test sets}
    \label{fig:det_curves_testsets}
\end{figure*}

\begin{figure*}[hbt]
    \centering
    \includegraphics[width=\textwidth]{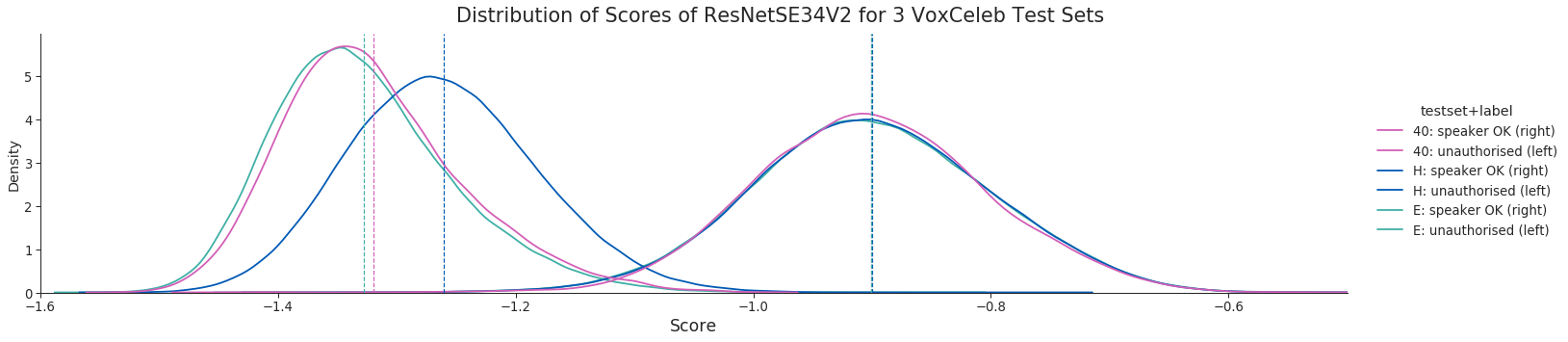}
    \caption{Distribution of scores for ResNetSE34V2 for 3 VoxCeleb test sets}
    \label{fig:scores_testsets}
\end{figure*}

Figures \ref{fig:det_curves_testsets} and \ref{fig:scores_testsets} show the overall DET curves and score distributions for the 3 test sets. From the DET curves it is evident that ResNetSE34V2 performs worst on VoxCeleb1-H. Performance for VoxCeleb1 test (40 speakers) and VoxCeleb1-E looks similar in the range of the operating threshold. However, the DET curve for the VoxCeleb1 test (40 speakers) is jagged and of poor quality due to the small sample size of the dataset. In Figure \ref{fig:scores_testsets} the score distributions for authorised speakers (right) are almost identical, but the score distriubtions of unauthorised speakers (left) are not. The distribution of the VoxCeleb1 test (40 speakers) and VoxCeleb1-E are left skewed, while VoxCeleb1-H has much greater kurtosis than the other two test sets. It is thus no surprise that VoxCeleb1-H, which contains appropriate speaker pairs to test realistic false positive scenarios, has a much higher FPR than VoxCeleb1 test (40 speakers) and -E, for the same FNR. Given these observations, application developers should consider VoxCeleb1-E as the easy test set, and VoxCeleb1-H as the heterogeneous test set. Of the 3 test sets, VoxCeleb1-H is the only appropriate test set for evaluating real-life embedded applications. Application developers should resist the urge of using an inappropriate test set to inflate the performance of their model.
    
The minimum value of the detection cost function $C_{Det}(\theta)$ is a popular metric to compare model performance in the speaker verification research community. Developers should nonetheless pay attention when using it to evaluate model performance for particular applications. Firstly, the parameters of the detection cost function should be selected so that the error rate weighting reflects the requirements of the application. Secondly, $C_{Det}(\theta)$ only presents a snapshot of model performance at a single threshold value. This value is generally not optimised for individual subgroups. DET curves provide a more holistic view of the performance of speaker verification models across subgroups and thresholds. They have been recommended by standards associations \cite{greenberg2020two}, yet many published papers do not show DET curves for their models. With \textbf{SVEva Fair} developers and researchers will be able to quickly produce DET curves for analysis, and thus present a more robust evaluation and comparison of speaker verification models.

\begin{mdframed}[style=greybox, frametitle={Take-away}]
    Real-life embedded speaker verification applications should carefully consider the choice of test set and error metrics to present robust evaluations and comparison across models. Of the test sets considered in this study, VoxCeleb1-H is the only test set with appropriate speaker pairs for evaluating realistic false positive scenarios that arise in embedded speaker verification applications. DET curves should be consulted when evaluating model performance.
\end{mdframed}

\subsection{Integrating SVEva Fair into the Embedded ML Development Pipeline}
\label{recommendations}
\begin{figure*}[hbt]
    \centering
    \subfigure[]{
        \includegraphics[width=0.18\textwidth]{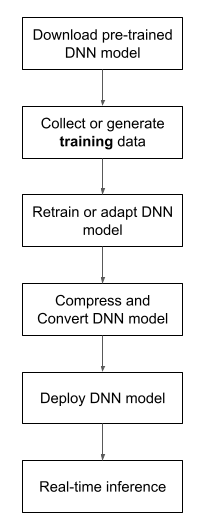}
        \label{fig:standard_workflow}
    }
    \subfigure[]{
        \includegraphics[width=0.75\textwidth]{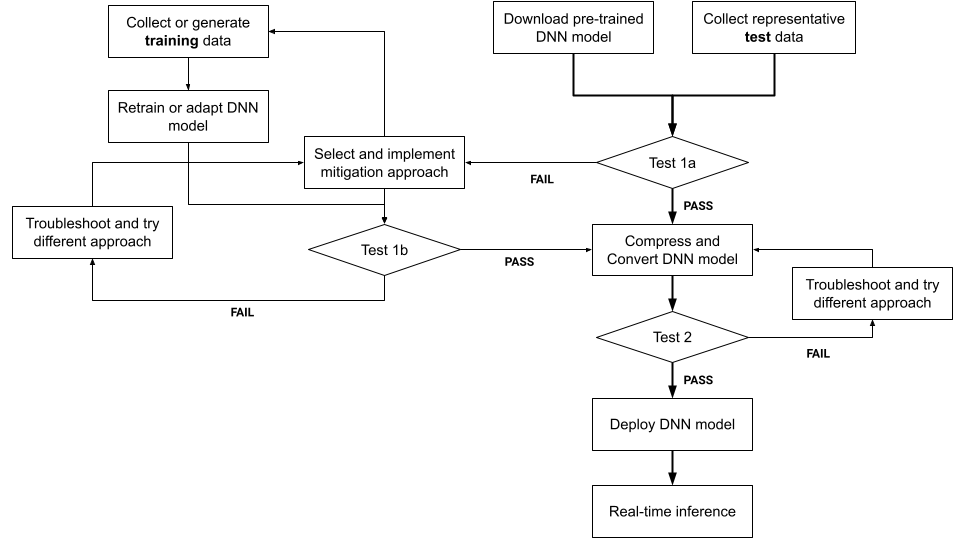}
        \label{fig:testing_workflow}
    }
    \caption{Embedded ML development workflow without (a) and with (b) fairness testing}
\end{figure*}

In the standard embedded ML development pipeline shown in Figure \ref{fig:standard_workflow}, a pre-trained speaker verification model is retrained or adapted with additional training data before it is compressed and converted to be deployed to embedded devices for real-time inference. \textbf{SVEva Fair} can be integrated into this workflow to ensure that the speaker verification component is fair, and that it retains its fairness as it undergoes different processing steps. Using \textbf{SVEva Fair} after model download as suggested with Test 1a in Figure \ref{fig:testing_workflow} enables developers to establish a fairness baseline using application specific test data. Testing model fairness at this stage has the advantage that initially only small amounts of test data need to be collected, and additional training data that is collected or generated can take the outcome of the test into account. This conserves and focuses resources in the development process. Moreover, it provides an opportunity for developers to consider a variety of mitigation strategies to improve model fairness. After a mitigation approach has been selected and implemented, Test 1b is suggested to evaluate if the fairness of the speaker verification component has improved sufficiently over the baseline. If the test passes, the development process can continue with compression and model conversion. If Test 1b fails, the mitigation approach should be revised. We suggest one further test after compression and conversion, to ensure that these processing steps do not produce surprising fairness challenges. After passing this final test, the fair speaker verification component is ready to be deployed for real-time inference.

\begin{mdframed}[style=greybox, frametitle={Take-away}]
        \textbf{SVEva Fair} can be integrated into the standard embedded ML development pipeline to test the fairness of speaker verification components after various processing steps. This enables developers to identify fairness challenges early on, to consider different strategies for mitigating fairness challenges and to focus their resources on meaningful actions that improve fairness.
\end{mdframed}

\subsection{Pragmatic Considerations}
\textbf{SVEva Fair} and the case study with VoxCeleb Trainer have some limitations. VoxCeleb is a celebrity dataset that is not representative of the broad public. Different models, or models trained on data representative of particular application contexts, may have different fairness characteristics. There is also a distribution shift between the VoxCeleb train and test sets. We did not attempt to retrain or adapt the available models to improve fairness, as this was out of scope. Given that speaker verification models should generalise to new speakers, and that VoxCeleb1 is frequently used to evaluate models trained on VoxCeleb2, we consider this a reasonable design choice. Moreover, absolute performance was not important to us, as the case study served the purpose of demonstrating how \textbf{SVEva Fair} supports the evaluation of relative performance differences between subgroups and models. As the language of the dataset is not specified, it is unclear whether nationality should be used as a proxy for accent, language or both. For our purpose, this does not affect how we apply \textbf{SVEva Fair} in our case study, as demographic subgroups can be defined flexibly. 

We intentionally inherited the demographic subgroups defined in the VoxCeleb1-H test set. The authors of the test set included all subgroups with 5 or more speakers. This means that some of the subgroups in our case study have very small sample sizes. We did not vary our analysis techniques for these subgroups, and some of the fairness results may be attributed to an insufficient sample size. However, the overarching observations still hold true. For example, performance differences between female and male speakers exist not only in subgroups with small samples, but also for speakers from the three nationalities most represented in the test set: the USA, UK and Canada. Subgroups with small sample sizes were also found amongst the top, mid and bottom performing subgroups. To generate reliable test results for subgroups a sufficient sample size should be selected based on statistical guidelines, rather than data availability. While the sample size of the test set affects the input to \textbf{SVEva Fair} and thus the validity of results, it does not change the evaluation mechanism itself.

Finally, the Fairness Index on its own does not provide a complete view of fairness across models, as it does not show how fairness varies across subgroups. Even though the index is dimensionless, it also has no inherent meaning that makes it possible to define sufficient or insufficient fairness for a model. For deeper model comparison it is thus necessary to consult the subgroup $C_{Det}\ ratios$ or the scatter plot visualisations that \textbf{SVEva Fair}r supports. While we have developed this framework to support the development of embedded speaker verification components, we have evaluated \textbf{SVEva Fair} on publicly accessible speaker verification models that were not specifically developed for embedded applications.

\section{Conclusion and Outlook}

Despite fairness being a major area of focus of traditional machine learning (ML), it is only an emerging consideration in embedded ML. Many open research directions exist to evaluate the fairness of existing techniques in embedded ML and TinyML, and to develop fair approaches that also retain their fairness in distributed, resource constrained and context-dependent applications.


In this study we have developed \textbf{SVEva Fair}, a framework for evaluating the fairness of speaker verification components. To our knowledge this is the first evaluation framework of its kind for embedded speaker verification applications. \textbf{SVEva Fair} successfully supports developers in two tasks: interrogating whether speaker verification model performance varies across subgroups, and comparing fairness across models. We present a detailed case study in which we use \textbf{SVEva Fair}r to evaluate and compare the fairness of two models released with the VoxCeleb Trainer benchmark, and trained on the VoxCeleb2 dataset. Using the evaluation measures and visualisations supported by \textbf{SVEva Fair}, we test these publicly accessible models on the VoxCeleb1-H dataset. Our evaluation shows that both models are not fair and perform significantly worse for female speakers of all but one nationality. Model performance also varies across nationalities. Interestingly, even though the overall performance varies between the two models, they have similar Fairness Indices. However, fairness varies inconsistently across subgroups across the models. Based on this work we present three key insights for speaker verification application developers. Firstly, speaker verification components should not be assumed to be fair, unless they have been tested for fairness for relevant subgroups. Secondly, existing best practices for evaluating speaker verification models are a useful starting point for evaluating fairness. To this end \textbf{SVEva Fair} leverages DET curves and the detection cost function $C_{Det}(\theta)$ to support developers in evaluating model performance across subgroups and comparing fairness across models. Finally, evaluating speaker verification fairness should be a part of the embedded ML workflow, and \textbf{SVEva Fair} can be integrated into the development pipeline.

As outlook and future work, we intend to apply \textbf{SVEva Fair} to evaluate the effect of model compression on speaker verification components. For embedded speaker verification, fairness challenges may also be amplified by hardware components. For example low quality microphones in low cost smartphones may increase the fairness challenges already experienced by demographic groups that use those devices. This kind of system-related bias is yet to be considered in the fairness community and is an open area for future research. One way to approach this is to consider the connection between reliability and fairness of embedded ML components. We promote viewing fairness challenges as an important category of reliability concern for edge intelligence. In fairness-induced reliability concerns the overall functioning or failure of embedded ML components is determined by a user's demographic attributes. Reliability has been well studied in software engineering, is considered an important aspect of trustworthiness in cyber-physical systems and has given rise to a reliability engineering discipline that considers mechanical component design and failure analysis. We believe that these disciplines will have valuable insights to offer to reliable and fair embedded speaker verification and edge intelligence more broadly.

\begin{acks}
This research was partially supported by TAILOR, a project funded by EU Horizon 2020 research and innovation programme under GA No 952215, and the iSafe project funded by TU Delft Safety \& Security Institute.
\end{acks}

\bibliographystyle{ACM-Reference-Format}
\bibliography{mendeley}

\appendix

\end{document}